\documentclass[aps,preprint,showpacs,amsmath,floatfix]{revtex4}
\usepackage{graphicx}%
\usepackage{dcolumn}
\usepackage{color}
\usepackage{amsmath}
\usepackage{here}

\usepackage{graphicx}

\begin{document}
\newcommand{\be}{\begin{equation}}
\newcommand{\ee}{\end{equation}}
\newcommand{\rojo}[1]{\textcolor{red}{#1}}

\title{Two-color discrete localized modes and resonant scattering\\
in arrays of nonlinear quadratic optical waveguides}

\author{Mario I.  Molina$^1$, Rodrigo A. Vicencio$^{1,2}$, and Yuri S. Kivshar$^3$}

\affiliation{$^1$Departamento de F\'{\i}sica, Facultad de
Ciencias, Universidad de Chile, Casilla 653, Santiago, Chile\\
$^2$Max-Planck-Insitut f\"ur Physik komplexer Systeme,
N\"othnitzerstr. 38, Dresden 01187, Germany\\
$^3$Nonlinear Physics Center, Research School of Physical Sciences
and Engineering, Australian National University, Canberra ACT
0200, Australia}

\begin{abstract}
We analyze the properties and stability of two-color discrete
localized modes in arrays of channel waveguides where tunable
quadratic nonlinearity is introduced as a nonlinear defect by
periodic poling of a single waveguide in the array. We show that,
depending on the value of the phase mismatch and the input power,
such two-color defect modes can be realized in three different
localized states. We also study resonant light scattering in the
arrays with the defect waveguide.
\end{abstract}


\maketitle

\section{Introduction}

The study of nonlinear propagation of light in periodic photonic
structures recently attracted strong interest due to the unique
possibility of observing experimentally an interplay between the
effects of nonlinearity and periodicity, including the generation
of {\em spatial optical solitons}~\cite{book}. Recently fabricated
nonlinear periodic structures such as arrays of weakly coupled
nonlinear optical waveguides can simultaneously support
distinctive types of self-trapped optical beams because linear
effects such as diffraction may differ dramatically compared to
those in the corresponding continuous systems~[2 - 5].

In addition to the nonlinear periodic systems, during last years a
growing interest is observed in the study of nonlinear optics
associated with the so-called {\em quadratic nonlinearities} which
may produce the effects resembling those known to occur in cubic
nonlinear materials. Typical examples are all-optical switching
phenomena in interferometric or coupler configurations as well as
the formation of spatial and temporal solitons in planar
waveguides~\cite{buryak}. One of the recent highlights in this
field is the first experimental demonstration of {\em discrete
solitons} with two frequency components mutually locked by a
quadratic nonlinearity~\cite{exp_chi2}. These optical experiments
have been performed in arrays of weakly coupled channel waveguides
with tunable cascaded quadratic nonlinearity, and they
demonstrated a good agreement with the theoretical
analysis~\cite{lederer}. As a matter of fact, it was demonstrated
that arrays of coupled channel waveguides fabricated in a
periodically poled Lithium Niobate slab represent a convenient
system to verify experimentally many theoretical predictions.
These experimental observations open many novel perspectives for
employing larger nonlinearities in the lattice systems with
quadratic materials.

In this paper, we analyze the properties and stability of
two-color discrete localized modes in arrays of channel
waveguides. In particular, we show that when periodic poling is
applied to just a single waveguide in the array, it creates a
nonlinear defect~\cite{defect,chi2_defect} that may support
strongly localized discrete modes similar to discrete solitons and
also display specific resonant scattering properties.

The paper is organized as follows. In Sec.~II we introduce our
model and find the profiles of two-color discrete localized modes.
We show that such modes can exist in three different states, and
they can be observed in the same array depending on the array
parameters and the value of the input power. Section~III is
devoted to the study of the resonant light transmission through
the defect waveguide that can be associated with the Fano
resonance. Finally, Sec.~IV concludes the paper.

\section{Two-color localized modes}

Being motivated by the design of the periodic photonic structures
recently employed for experiments~\cite{exp_chi2}, we consider an
array of weakly coupled linear waveguides where one waveguide has
periodic poling, and therefore it possesses a quadratic nonlinear
response. When the matching conditions are satisfied, the
fundamental-frequency (FF) mode with the frequency $\omega$
generates the second-harmonic (SH) wave at the frequency
$2\omega$, so that such a structure with the poled waveguides may
behave as a nonlinear defect with localized quadratic
nonlinearity.

In the tight-binding approximation~\cite{review2}, the effective
equations for the complex envelopes of the FF wave ($u_n$) and its
SH component ($v_n$) coupled at the defect waveguide (at $n=0$)
can be written in the form
\begin{equation}  \label{eq:1}
   \begin{array}{l} {\displaystyle
      i {du_{n}\over{d z}} + c_{u}( u_{n+1} + u_{n-1} ) + 2 u_{0}^{*} v_{0} \delta_{n 0} =0,
   } \\*[9pt] {\displaystyle i {dv_{n}\over{d z}} + c_{v}( v_{n+1} + v_{n-1} ) -\Delta v_{n} +
u_{0}^{2} \delta_{n 0}=0, }
\end{array}
\end{equation}
where $c_u$ and $c_v$ are the coupling coefficients, $\Delta$ is
the phase mismatch parameter, and $\delta_{n0}$ is a
delta-function.

The discrete model (\ref{eq:1}) has two conserved quantities, the
total power
\be
P = \sum_{n = -\infty}^{\infty} (|u_{n}|^{2} + 2|v_{n}|^{2})
\label{eq:power}
\ee
and the Hamiltonian. We look for stationary solutions of the
coupled equations in the form
\[
u_{n}(z) = u_{n} e^{i \lambda z}, \;\;\; v_{n}(z) = v_{n} e^{2i
\lambda z},
\]
where we assume, without loss of generality, that the amplitudes
$u_n$ and $v_n$ are real.  As a result, Eqs.~(\ref{eq:1}) become
\begin{equation}  \label{eq:5}
\begin{array}{l} {\displaystyle
\lambda u_{n} = c_{u}(u_{n+1} + u_{n-1}) + 2u_{0} v_{0}\ \delta_{n
0}, } \\*[9pt] {\displaystyle 2 \lambda v_{n} = c_{v}(v_{n+1} +
v_{n-1}) -\Delta\ v_{n} + u_{0}^{2}\ \delta_{n 0}. }
\end{array}
\end{equation}

We look for spatially localized discrete states  in the form
$u_{n} = u_{0} \eta^{|n|}$ and $v_{n} = v_{0} \xi^{|n|}$, where
$|\eta|, |\xi| <1$, and obtain from Eqs.~(\ref{eq:5}) at $n \neq
0$ two relations, $\lambda = c_{u}(\eta + \eta^{-1})$ and
$2\lambda = c_{v}(\xi + \xi^{-1}) - \Delta$, which can be combined
to yield
\be c_{v}\left(\xi + {1\over{\xi}}\right)-2 c_{u}\left(\eta +
{1\over{\eta}}\right) =\Delta.
\label{F} \ee
On the other hand, for the defect waveguide at $n = 0$ we obtain
$\lambda = 2 c_{u} \eta + 2 v_{0}$ and $2\lambda v_{0} = 2 c_{v}
v_{0} \xi - \Delta v_{0} + u_{0}^{2}$.
After some simple algebra, we solve for the parameters $u_{0}$ and
$v_{0}$ in terms of $\eta$ and $\xi$ and obtain
\be v_{0} = {c_{u}\over{2}}\left( {1\over{\eta}} - \eta \right),
\;\;\;\; u_{0}^{2} = c_{v} v_{0} \left( {1\over{\xi}} -
\xi\right).\label{eq:2ppp}
\ee
In addition, Eq.~(\ref{eq:power}) can be expressed as
\be P= u_{0}^{2}\ \left( {1 + \eta^{2}\over{1 - \eta^{2}}}\right)
+ 2\ v_{0}^{2}\ \left( {1 +  \xi^{2}\over{1 - \xi^{2}}} \right),
\label{eq:power2} \ee
and the next step is to replace the amplitudes $u_0$ and $v_0$ in
Eq.~(\ref{eq:power2}) and obtain the second equation,
\be c_{u} c_{v} \frac{(1 - \xi^{2})}{2\eta \xi}(1 + \eta^{2}) +
c_{u}^{2} \frac{(1 - \eta^{2})^2}{2\eta^2} \frac{(1 + \xi^{2})}{(1
- \xi^{2})}=P. \label{G} \ee
Equations (\ref{F}) and (\ref{G}) constitute a system of two
coupled nonlinear equations for the functions $\eta$ and $\xi$.
Moreover, Eq.~(\ref{F}) depends on the mismatch parameter
$\Delta$, whereas Eq.~(\ref{G}) depends on the total power $P$.

First, we assume that $c_{u}, c_{v} > 0$. From Eq.~(\ref{eq:2ppp})
we find that if $v_{0} > 0$ then $0 < \xi, \eta < 1$. On the other
hand, if $v_{0} < 0$, then $-1 < \xi, \eta < 0$. Also, without
loss of generality, we can consider only $u_{0}>0$. Thus, we
should find real solutions for the values $\eta$ and $\xi$ in the
domain defined by the conditions $0 < \eta, \xi <1$ and $-1 <
\eta, \xi < 0$ on the plane ($\eta, \xi$).

From the structure of Eqs.~(\ref{F}) and (\ref{G}) we notice that
a change $\Delta \rightarrow -\Delta$ is equivalent to the change
$(\eta, \xi) \rightarrow (-\eta, -\xi)$. Thus, we only need to
consider positive values of $\Delta$. In order to find spatially
localized states, we should analyze the functions of $\eta$ and
$\xi$ defined by the left-hand sides of the expressions (\ref{F})
and (\ref{G}), for varying values of the mismatch parameter
$\Delta$ and the total power $P$. This analysis reveals that, in
general, for a given value of $\Delta$, a minimum value of the
power is required to support a localized state at the defect.

\begin{figure}[tbp]
\includegraphics[width=3.0in,height=62mm]{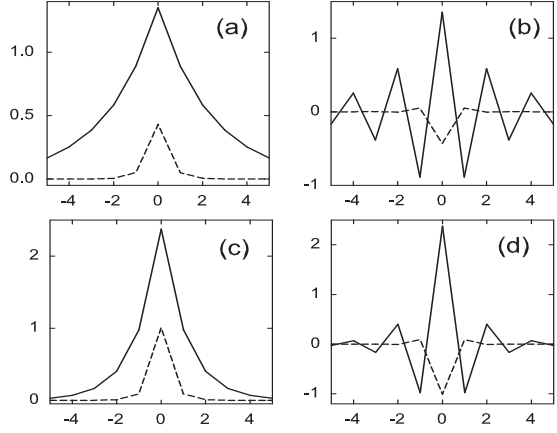}
\caption{Examples of localized modes at $\Delta = 0$ and two
values of the total power above the threshold: (a,b) $P=5$ and
(c,d) $P=10$. Solid and dashed curves show the
fundamental-frequency ($u_n$) and second-harmonic ($v_n$) fields ,
respectively. } \label{fig1}
\end{figure}

As an example, in Figs.~1(a-d) we show several profiles of the
localized modes for the case $\Delta =0$ and two values of the
power $P$, for the typical values $c_{u}=1$ and $c_{v} = 0.5$.
Below a certain threshold power, there exist no localized states,
but above the threshold two localized modes appear, unstaggered
(a) and staggered (b) modes,  respectively [see Figs.~1(a,b)]. As
the total power $P$ increases, these two modes become more
localized, as shown in Figs.~1(c,d).

\begin{figure}[tbp]
\includegraphics[width=3.3in,height=70mm]{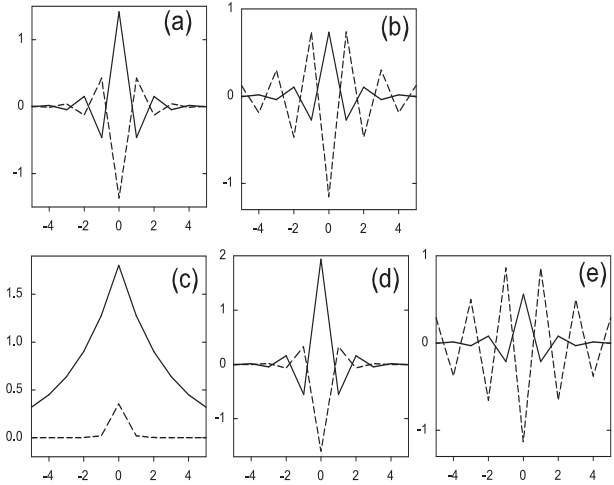}
\caption{Same as in Fig.~1 but for $\Delta = 5$ and two values of
the total power above the threshold: (a,b) $P=7$ and (c-e) $P=10$.
In this case, two staggered localized modes appear about the
threshold power, and then additional unstaggered mode appears for
larger values of $P$. } \label{fig2}
\end{figure}

For a finite positive mismatch $\Delta$ and increasing power $P$,
we find that there is a threshold power below which no localized
states exist. When the power exceeds the first threshold, at least
one (staggered) localized modes becomes possible and then the
other mode appears for larger powers. As the value of the mismatch
$\Delta$ increases further, it can be proven that the contour
curve associated with Eq.~(\ref{F}), which depends on $\Delta$,
experiences a curvature change at $\Delta = 2(2 c_{u} - c_{v}) =
3$ and, as a result, a double root is possible at a certain power
level. Further power increase gives rise to two staggered
localized states. Even further increase in the power makes
possible another (unstaggered) localized state to exist, as shown
in Fig.~2 at $\Delta = 5$.

The threshold value of the total power that corresponds to the
appearance of the first localized mode can be found analytically
in the form,
\be P_{\rm min}^1(\Delta) = c_{u} [(\Delta + 4 c_{u})^{2} - 4
c_{v}^{2}]^{1/2}, \ee
for the condition $\Delta > 2(c_{v} - 2 c_{u})$, and
\be P_{\rm min}^2(\Delta) = c_{u} [(\Delta - c_{u})^{2} - 4
c_{v}^{2}]^{1/2}, \ee
for the condition $\Delta < 2(2 c_{u} - c_{v})$.  In the first
case, the lowest localized mode is unstaggered, whereas in the
second case, it is staggered. Applying these results to Fig.~1, we
obtain $P_{\rm min}^1(0) = P_{\rm min}^2(0) = \sqrt{15} = 3.873$.

By solving Eq.~(\ref{eq:1}) numerically, we have checked the
stability of all localized modes to propagation, and we came to
the conclusion that such modes are stable, and they do not display
any bistability.

\section{Resonant transmission}

The two-color localized modes described above can be excited and
generated experimentally in arrays of weakly coupled quadratic
waveguides being detected through the specific features of the
transmission coefficient~\cite{OL_fano}. Therefore, here we
analyze the scattering of a plane FF wave by a quadratic defect
waveguide. When the phase-matching conditions are satisfied, after
the interaction with the quadratic waveguide, the FF wave
generates a SH wave which could either propagate or get trapped
being guided by the defect waveguide in the form of a localized
mode.

{\em Propagating SH field.} To calculate the transmission
coefficient in the case of the propagating SH field, we present
the fields as
\be u_{n}(z) = e^{i \lambda_u z} \left\{ \begin{array}{ll}
    a e^{i k n} + b_1 e^{-i k n}, & \mbox{$n < 0$}\\
    c_1 e^{i k n},   & \mbox{$n \geq 0$},
    \end{array}
        \right.
\ee
\be \label{eq:vn}
v_{n}(z) = e^{i \lambda_v z} \left\{
\begin{array}{ll}
    b_2 e^{-i q n}, & \mbox{$n < 0$}\\
    c_2 e^{i q n}, & \mbox{$n \geq 0$},
    \end{array}
        \right.
\ee
where $a$ is the amplitude of the FF wave before the scattering,
$b_{1,2}$ and $c_{1,2}$ are the amplitudes of the reflected and
transmitted FF and SH waves, respectively, and the wave numbers
$k$ and $q$ are defined in the domain $0 < k, q < \pi$. Using
these expressions far from the defect site $n=0$, from the
phase-matching condition $\lambda_v = 2 \lambda_u$ we obtain
\be c_{v} \cos \, q - \frac{\Delta}{2} = 2 c_{u} \cos \, k.
\label{eq:vu} \ee
Equation (\ref{eq:vu}) defines $q = q(k)$ which is real and
positive in some interval, $k_{\rm min} < k < k_{\rm max}$, where
\[
k_{\rm min, max} = \cos^{-1}\left(\frac{\Delta \pm 2c_{v}}{4
c_{u}}\right).
\]
Outside this interval, $q$ is purely imaginary, and this case
corresponds to the SH field localized at the waveguide.

Evaluating the fields at the sites $n=-1,0$, we obtain
\begin{equation}  \label{eq:A}
   \begin{array}{l} {\displaystyle
\lambda_{u} c_1 = c_{u} (c_1 e^{i k} + a e^{-i k} + b_1 e^{i k}) +
2c_1^*c_2,
   } \\*[9pt] {\displaystyle \lambda_{v} c_2 = c_{v} (c_2 e^{i q} + b_2 e^{i q}) -c_2 \Delta +
c_1^2, }
\end{array}
\end{equation}
\begin{equation}
c_1 =a + b_1, \;\;\; b_2 = c_2,
\label{eq:D}
\end{equation}
the latter condition implies that the SH fields are generated
symmetrically. Using these results, we find the important relation
$c_2 = (i/2 c_{v} \sin \,q) c_1^{2}$ that allows us to obtain the
nonlinear equation for the transmission coefficient of the FF
wave, defined as $t(k)= |c_1|^2/|a|^2$,
\be t(k)= {1\over{[1 + B(k) t(k)]^{2} }}, \label{eq:t}
\ee
where $B(k) \equiv |a|^2 A(k) = |a|^2/\{2 c_{u} c_{v} \sin k \,
\sin q(k)\}$. By rewriting Eq.~(\ref{eq:t}) in the form
\begin{equation}
\label{cubic} A^2(k) t^{3} +2 B(k) t^{2}+ t = 1,
\end{equation}
and using the fact that $A(k) > 0$, it is easy to see that this
result does not predict bistability. As a matter of fact, the only
real and positive solution of Eq.~(\ref{eq:t}) can be found in an
analytical form from the equation (\ref{cubic}).

{\em Localized SH field.} To calculate the transmission
coefficient in the case of the localized SH field, we present the
SH field as
\be v_{n} = e^{i \lambda_v z} \left\{
\begin{array}{ll}
    b_2 e^{q n}, & \mbox{$n \leq 0$},\\
    c_2 e^{-q n},  & \mbox{$n \geq 0$},
\end{array}
\right. \ee
which can be obtained by replacing $q\rightarrow i q$ in
Eqs.~(\ref{eq:vn}). Real and positive values of $q$ can be found
from Eq.~(\ref{eq:vu}) that now takes the form:
\be c_{v} \cosh q - \frac{\Delta}{2} = 2 c_{u} \cos k. \ee
Wavenumber $k$ of the incident FF field should be taken outside of
the domain $(k_{\rm min}, k_{\rm max})$, i.e. for $0 < k < k_{\rm
min}$ and $k_{\rm max} < k < \pi$. The relation between the
transmitted amplitudes now becomes $c_2 = c_1^2/(2 c_v \sinh q)$,
and the transmission coefficient $t(k)$ is defined by the equation
\be t(k) = \frac{1}{1 + D^2(k) t^{2}(k)},
\label{eq:32}
\ee
where $D(k) \equiv |a|^2/[2c_vc_u \sin k \sinh q(k)]$. Equation
(\ref{eq:32}) possesses a real and positive solution for $t(k)$
that can be found in an analytic form.

Figures 3(a-d) show some examples of the dependence of the
transmission coefficient $t(k)$ on the wavenumber $k$ ($0 < k <
\pi$) for different values of the coupling coefficients $c_{u}$,
$c_{v}$ and the mismatch parameter $\Delta$. In a sharp contrast
with the problem of the cubic defect analyzed earlier in
Refs.~\cite{cubic} where the transmission coefficient does not
show any interesting structure and possesses a single maximum
only, here the transmission coefficient shows two minima
associated with the resonant suppression of the transmission. For
small values of the ratio $c_{v}/c_{u}$, the middle interval is
narrow and the transmission remain low. The effect of the nonzero
mismatch parameter [see Fig.~3(d)] is to shift the position of the
middle interval making the transmission curve asymmetric.

The resonant suppression of transmission at some points, i.e.
$t(k_{\rm min})=t(k_{\rm max})=0$, corresponds to a novel type of
the well-known Fano resonance~\cite{flach} as recently discussed
in Ref.~\cite{OL_fano} for a simpler model. Indeed, destructive
interference and resonant suppression of transmission is observed
when there exists an extra localized state coupled to the
propagating mode with the energy inside the linear spectrum.
Indeed, the values $q(k_{\rm min})=0$ and $q(k_{\rm max})=\pi$
define the band edges of the propagation spectrum of the SH field,
and such resonances take place when the SH field is generated.

\begin{figure}[tbp]
\includegraphics[width=3.4in,height=65mm]{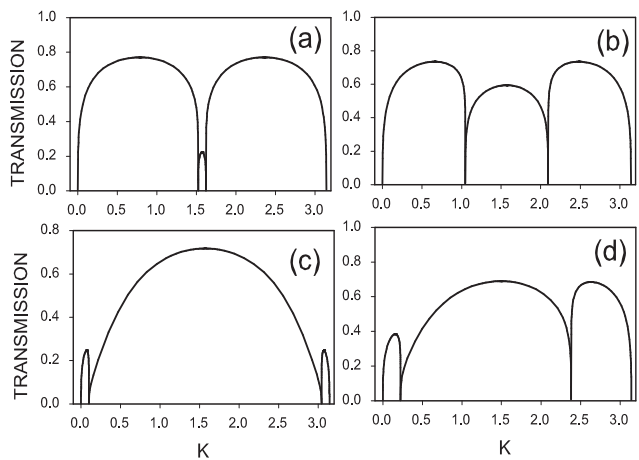}
\caption{Transmission coefficient of a plane wave (shown for the
fundamental frequency only) after scattering by a nonlinear
quadratic defect, for $\Delta =0$ and (a) $c_v=1$, $c_u=0.1$, (b)
$c_v=c_u=1$, and (c) $c_v=1$, $c_u=1.99$. The case (d) show the
asymmetric dependence for $\Delta =0.5$ and $c_v=1$, $c_u=1.7$. }
\label{fig3}
\end{figure}

\section{Conclusions}

Being driven by the recent successful experimental demonstrations
of discrete optical solitons with two frequency components
mutually locked by a quadratic nonlinearity, in this paper we have
studied two-color localized modes in arrays of channel waveguides.
We have assumed that the tunable quadratic nonlinearity is
introduced as a nonlinear defect by periodic poling of a single
waveguide in the array, and we have analyzed the structure and
stability of discrete localized modes created by mutual locking of
two frequency components. We have shown that, depending on the
value of the phase mismatch and the input power, such two-color
defect modes can be realized in the same array in three different
localized states, and we have studied also the resonant light
scattering in the array with the defect waveguide drawing
analogies with the Fano resonance.

\section{Acknowledgements}

The authors thank Andrey Miroshnichenko for useful discussions and
collaborations. This work was partially supported by Conicyt and
Fondecyt grants 1020139 and 7020139 in Chile, and by the
Australian Research Council in Australia.

\end{document}